\documentclass[twocolumn,showpacs,preprintnumbers,prl,amsmath,amssymb]{revtex4-1}

\usepackage[pdftex]{graphicx}
\usepackage{dcolumn}
\usepackage{bm}

\begin{document}

\title{Anisotropic and quasipropagating spin excitations in superconducting Ba(Fe$_{0.926}$Co$_{0.074}$)$_2$As$_2$}
\author{H.-F. Li,$^1$ C. Broholm,$^2$ D. Vaknin,$^1$ R. M. Fernandes,$^1$ D. L. Abernathy,$^3$ M. B. Stone,$^3$ D. K. Pratt,$^1$ W. Tian,$^1$ Y. Qiu,$^{4,5}$
N. Ni,$^1$ S. O. Diallo,$^1$ J. L. Zarestky,$^1$ S. L. Bud$'$ko,$^1$ P. C. Canfield,$^1$ and R. J. McQueeney$^1$}
\affiliation{$^1$Ames Laboratory and Department of Physics and Astronomy, Iowa State University, Ames, Iowa 50011, USA\\
$^2$Institute for Quantum Matter and Department of Physics and Astronomy, The Johns Hopkins University, Baltimore, Maryland 21218, USA\\
$^3$Oak Ridge National Laboratory, Oak Ridge, Tennessee 37831, USA\\
$^4$NIST Center for Neutron Research, National Institute of Standards and Technology, Gaithersburg, Maryland 20899, USA \\
$^5$Department of Materials Science and Engineering, University of Maryland, College Park, Maryland 20742, USA}
\date{\today}
\begin{abstract}
Inelastic neutron scattering from superconducting (SC) Ba(Fe$_{0.926}$Co$_{0.074}$)$_2$As$_2$ reveals anisotropic and quasi-two-dimensional (2D) magnetic
excitations close to \textbf{Q}$_{\texttt{AFM}}=(\frac{1}{2} \frac{1}{2})$ - the 2D antiferromagnetic (AFM) wave-vector of the parent BaFe$_2$As$_2$
compound. The correlation length anisotropy of these low energy fluctuations is consistent with spin nematic correlations in the $J_1$-$J_2$ model with
$J_1/J_2 \sim$ 1. The spin resonance at $\sim$8.3 meV in the SC state displays the same anisotropy. The anisotropic fluctuations experimentally evolve into
two distinct maxima only along the direction transverse to \textbf{Q}$_{\texttt{AFM}}$ above $\sim$80 meV indicating unusual quasi-propagating excitations.
\end{abstract}
 \maketitle

The unconventional superconductivity in the newly discovered \emph{R}FeAs(O$_{1-\texttt{x}}$F$_\texttt{x}$) (\emph{R} = rare earth) superconductors
\cite{Kamihara2008} with critical temperatures up to 55 K \cite{Chen2008,Ren2008} emerges upon suppression of an antiferromagnetic (AFM) phase and is
accompanied by a spin resonance in the magnetic excitation spectrum \cite{Christianson2008-1,Lumsden2009-1,Pratt2009,Chi2009,Li2009,Qiu2009}, indicating a
close connection between magnetism and superconductivity.

In the tetragonal structure of parent ferropnictides, degenerate magnetic states and frustration arise from competition between nearest-neighbor (NN) and
next-NN (NNN) AFM exchange couplings. Magnetic frustration is thought to produce an emergent nematic degree of freedom \cite{Chandra1990}, which couples to
orthorhombic distortions, inducing a structural transition \cite{Fang2008,Xu2008,Fernandes2009}. Nematic fluctuations may also be present in the
superconducting (SC) phase \cite{Fernandes2009}, potentially leading to local spin and electronic anisotropies within the Fe layers. The relationship
between nematicity and unconventional superconductivity in iron arsenides \cite{chuang2010} is a subject under intense debate.

Here we examine the in-plane wave-vector dependence of magnetic fluctuations in SC Ba(Fe$_{0.926}$Co$_{0.074}$)$_2$As$_2$ \cite{Ni2008-1} over a wide range
of energies using inelastic neutron scattering (INS). Our results show that the anisotropic spin fluctuations within the Fe layer consistent with the
square symmetry of the Fe layer evolve into unusual modes propagating \emph{only} along the direction of ferromagnetic (FM) near neighbor spin
correlations. We also find an analogous \textbf{Q}-space anisotropy of the spin resonance in the SC state.

High-quality Ba(Fe$_{0.926}$Co$_{0.074}$)$_2$As$_2$ single crystals display a sharp SC transition at $T_\texttt{c} \approx$ 22.2 K \cite{Ni2008-1}. INS
measurements were performed on the wide Angular-Range Chopper Spectrometer (ARCS) at the Spallation Neutron Source with incident energies up to 250 meV and
on the HB-3 triple-axis spectrometer at the High Flux Isotope Reactor with fixed final energy at 14.7 meV. The mosaic of the coaligned samples ($\sim$5.40
g, 26 crystals) for the ARCS (\emph{H}\emph{H}\emph{L}) measurements is $\sim$3.30$^\texttt{o}$ and $\sim$3.50$^\texttt{o}$ full width at half maximum for
rotations about the $(HH0)$ and $(00L)$ directions, respectively. The HB-3 measurements in the (\emph{H}\emph{K}0) scattering plane were carried out on a
realigned subset of six of the ARCS crystals with a total mass of $\sim$2.58 g and a mosaic spread of $\sim$0.53$^\texttt{o}$ for both $(H00)$ and $(0K0)$
directions.

\begin{figure}
\centering \includegraphics[width = 0.42\textwidth] {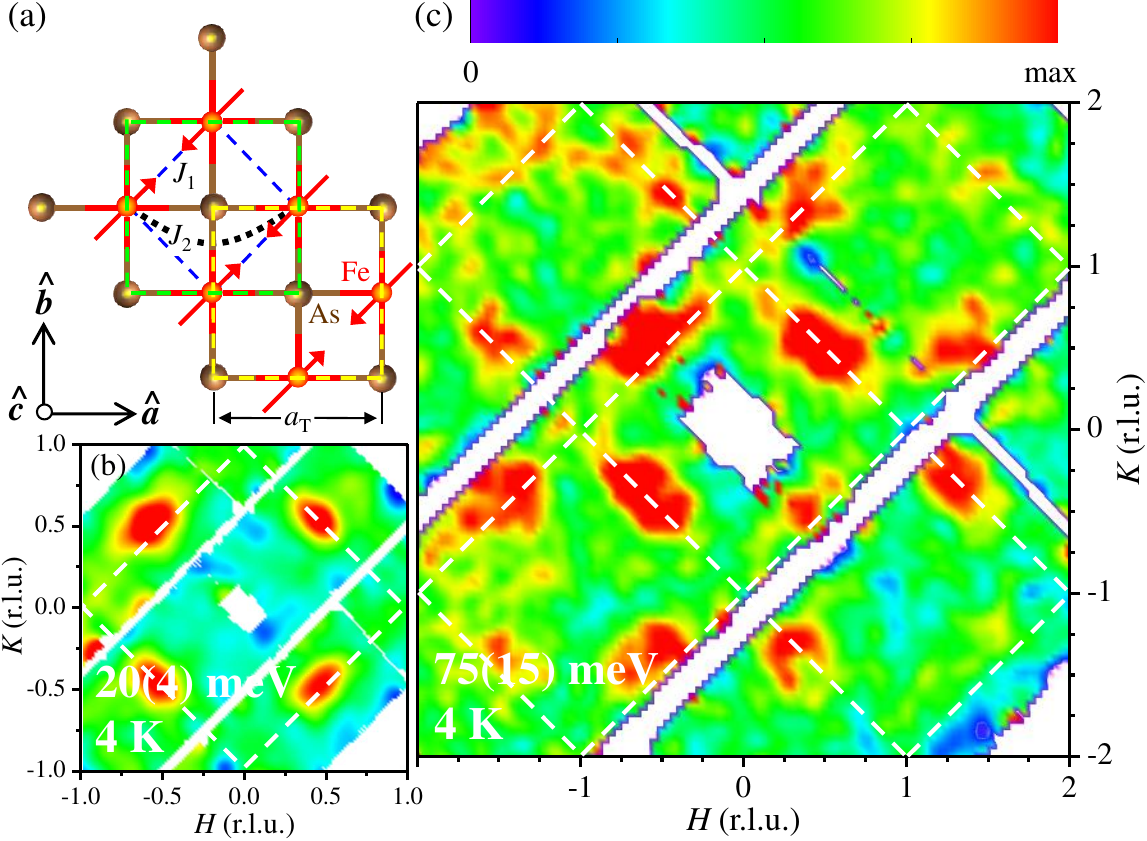}
\caption{(color online) (a) A single Fe layer showing a local spin configuration corresponding to $\textbf{Q}_{\texttt{AFM}} = (\frac{1}{2} \frac{1}{2})$ and
NN (NNN) exchange interactions $J_1$ ($J_2$). The tetragonal and Fe lattice unit cells are shown as dashed squares. Overview of the
${\bf Q}_{ab}={\bf Q}-({\bf Q}\cdot \hat{\bf c})\hat{\bf c}$ dependence at 4 K measured on ARCS with (b) $E_\texttt{i}$ = 50 meV and $\hbar\omega$ = 20 $\pm$ 4 meV
and (c) $E_\texttt{i}$ = 250 meV and $\hbar\omega$ = 75 $\pm$ 15 meV. An empty sample-holder background was subtracted. The dashed lines in [(b) and (c)] show
the Brillouin-zone boundary of the Fe square lattice.}
\label{Figure1-1}
\end{figure}

An overview of the $\bf Q$ dependence of the magnetic scattering at 4 K measured by the ARCS is shown in Figs.\ \ref{Figure1-1}(b) and \ref{Figure1-1}(c).
Due to the 2D nature of the spin fluctuations \cite{Christianson2008-1,Lumsden2009-1,Lester2010}, we present \emph{L}-integrated data. Maxima are observed
for $(H,K)=(m+\frac{1}{2},n+\frac{1}{2})$, $m$ and $n$ being integers, as for the AFM stripe ordering in BaFe$_2$As$_2$. The scattering peaks are quite
broad and feature a distinct anisotropy, indicating short-ranged spin correlations with an anisotropic correlation area in the paramagnetic state. Similar
observations in Ba(Fe$_{0.935}$Co$_{0.065}$)$_2$As$_2$ \cite{Lester2010} and paramagnetic CaFe$_2$As$_2$ \cite{Diallo2010} suggest that this anisotropy is
a universal property of spin fluctuations in the iron arsenides. The spin-space anisotropy of magnetic fluctuations in BaFe$_{1.9}$Ni$_{0.1}$As$_2$
\cite{Lipscombe2010} bears not directly related to the reciprocal space anisotropy observed here.

\begin{figure}
\centering \includegraphics[width = 0.42\textwidth] {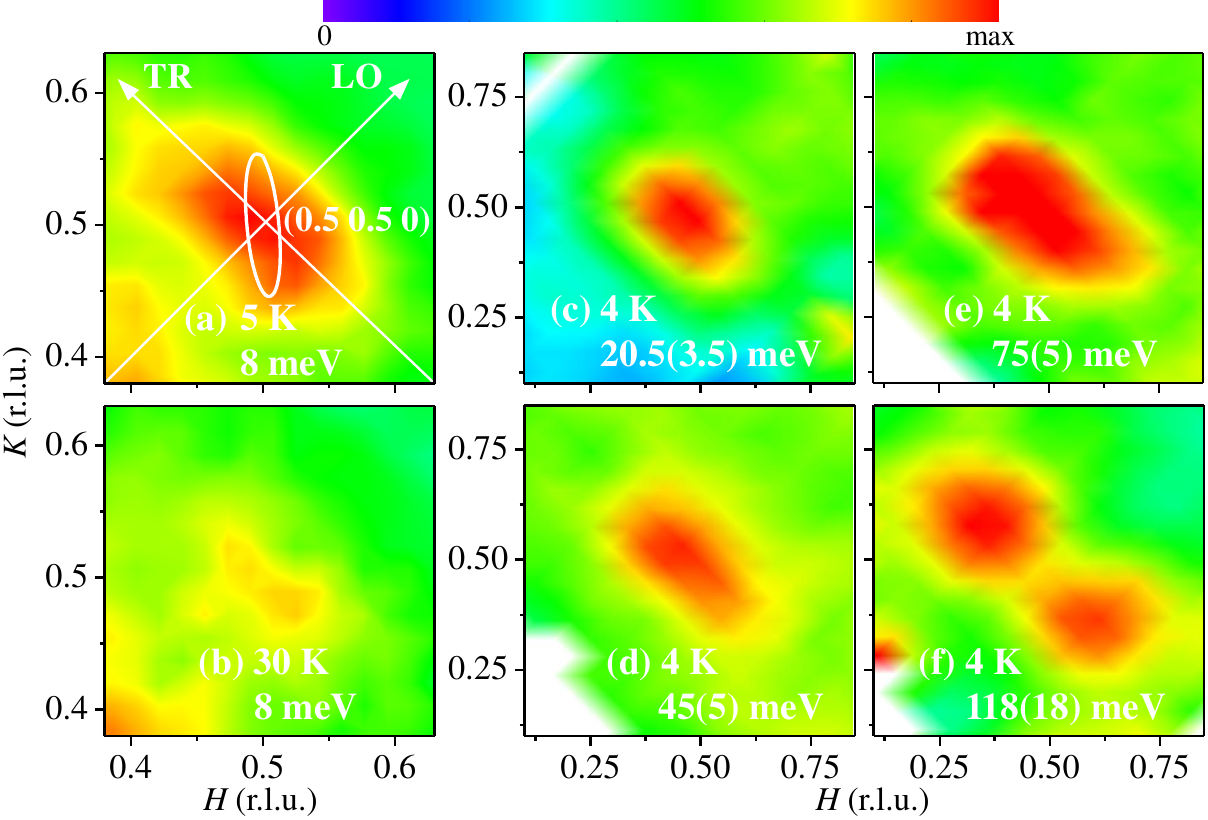}
\caption{(color online) Energy and wave-vector dependences around ${\bf Q}_{ab} = (\frac{1}{2} \frac{1}{2})$ [(a) and (c)-(f)] in the SC state and (b) in the
normal state. (a) and (b) show the ${\bf Q}=(HK0)$ dependence at the resonance energy $\sim$8 meV with the same intensity scale from HB-3. [(c)-(f)] Integrated data
over (c) 17-24 meV, (d) 40-50 meV,  (e) 70-80 meV, and (f) 100-136 meV with (c) $E_\texttt{i}=50$~meV and [(d)-(f)] 250 meV averaged over two quadrants from ARCS.
An empty sample-holder background was subtracted.}
\label{Figure2-2}
\end{figure}

In Figs. \ref{Figure2-2}(c)-\ref{Figure2-2}(f), all scattering patterns display a twofold symmetry with respect to $\textbf{Q}_{\texttt{AFM}}$. In
particular the {\bf Q}-width is considerably larger along the direction transverse (TR) to \textbf{ Q}$_{\texttt{AFM}}$ than along the longitudinal optical
(LO) direction. Two distinct maxima split off in the TR direction at $\sim$100 meV as shown in Fig.~\ref{Figure2-2}(f). Figure \ref{Figure1-1}(a) shows
that the TR direction at $(\frac{1}{2}\frac{1}{2})$ corresponds to FM spin correlations in the stripe AFM structure. At first sight, it might seem that
this twofold symmetry breaks the fourfold symmetry of the Fe square lattice. However, Fig. 1 shows that the $(\frac{1}{2} \frac{1}{2})$ point already has
twofold symmetry. While the twofold symmetry of the observed scattering pattern thus does \emph{not} break the symmetry of the Fe sublattice, it indicates
a spin-correlation area that is anisotropic with respect to the direction of correlated NN spins.

\begin{figure} \centering \includegraphics[width = 0.42\textwidth] {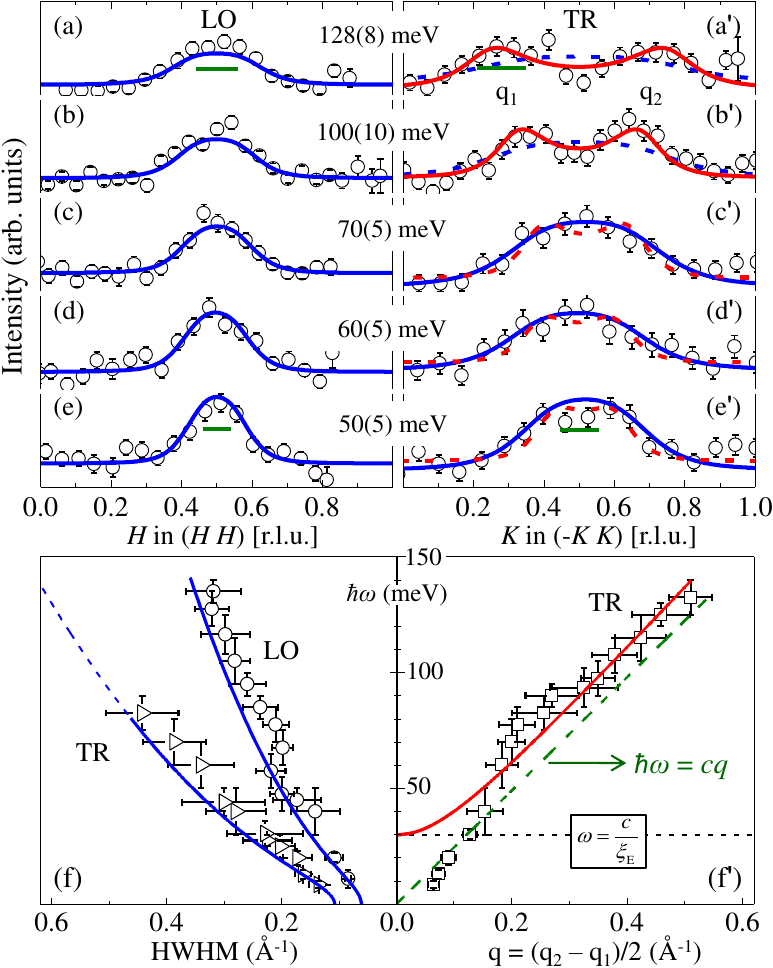}
\caption{(color online) [(a)-(e)] LO and [(a$'$)-(e$'$)] TR scans (circles) around ${\bf Q}_{ab} = (\frac{1}{2} \frac{1}{2})$ with $E_\texttt{i}$ = 250 meV
and energy transfers as indicated at 4 K from ARCS. The approximate \emph{L} (r.l.u.) ranges are [(a) and (a$'$)] [6.4, 7.5], [(b) and (b$'$)] [4.6, 5.8],
[(c) and (c$'$)] [3.2, 3.8], [(d) and (d$'$)] [2.7, 3.2], and [(e) and (e$'$)] [2.2, 2.7]. The solid lines are fits of [(a)-(e) and (c$'$)-(e$'$)] the diffusive
model and [(a$'$) and (b$'$)] the ballistic model. The dashed lines are calculated with [(a$'$) and (b$'$)] the diffusive model and [(c$'$)-(e$'$)] the ballistic
model with fitted parameters held fixed. The horizontal bars in (a), (a$'$), (e), and (e$'$) represent the expected \emph{q} resolutions. (f) compares the width
of single Lorentzian fits along LO (circles) and TR (triangles) directions to the HWHM of the diffusive model (lines). (f$'$) compares the position of TR
Lorentzian peaks (squares) to the dispersion (slanted dashed line) and peak splitting of the ballistic model (solid line). The horizontal dashed line corresponds
to the energy $c/\xi_\texttt{E}$ as described in the text.}
\label{Figure3-3}
\end{figure}

\begin{figure}
\centering \includegraphics[width = 0.42\textwidth] {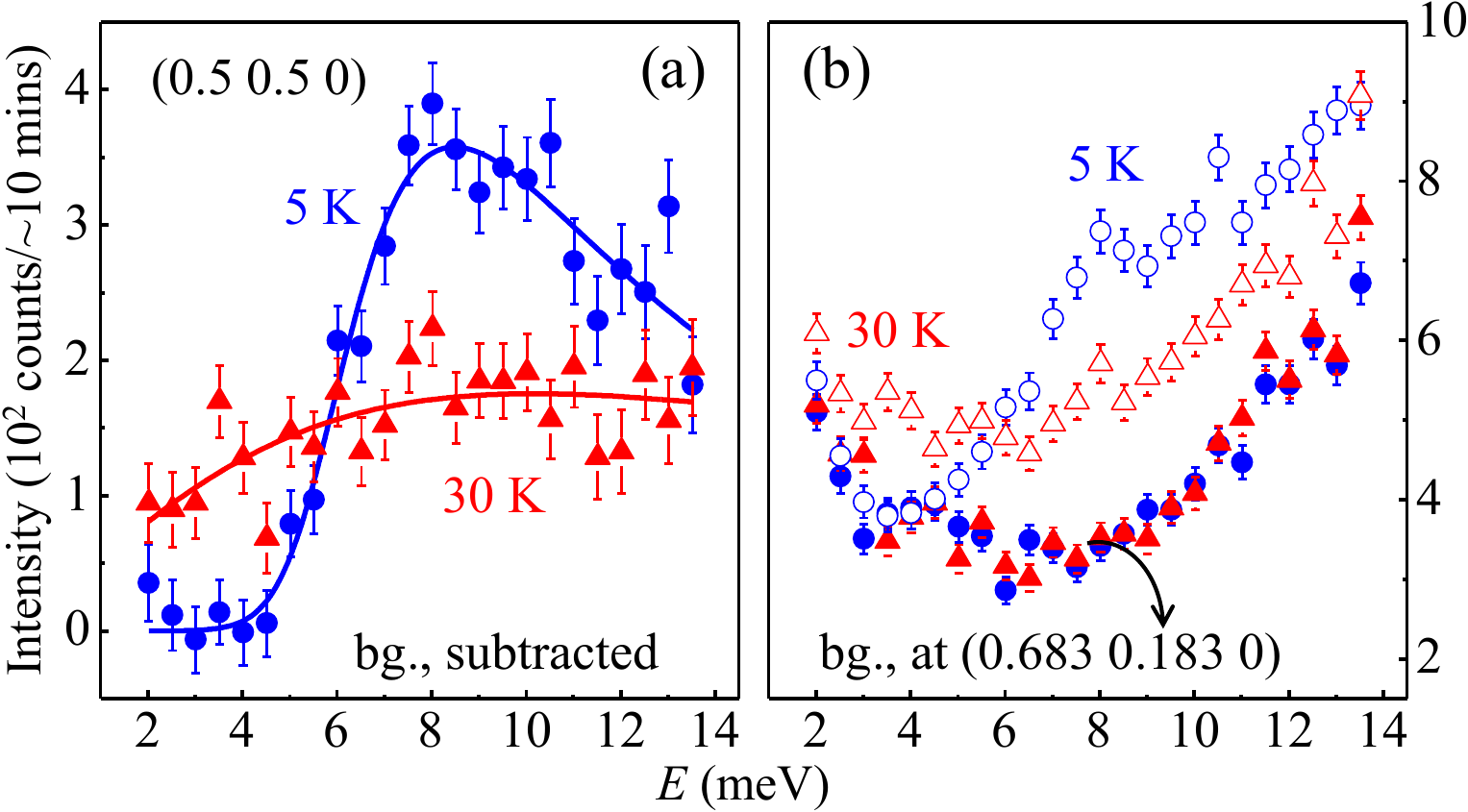}
\caption{(color online) (a) Energy dependence of INS at ${\bf Q}=(\frac{1}{2} \frac{1}{2} 0)$ below (5 K) and above (30 K) $T_\texttt{c}$
(solid symbols) after background subtraction. The solid lines are fits as described in the text. (b) Raw scattering data as well as their respective backgrounds
taken at \textbf{Q} = (0.683 0.183 0) from HB-3.}
\label{Figure4-4}
\end{figure}

The quality of the data can be ascertained in constant energy cuts passing through $(\frac{1}{2} \frac{1}{2})$ along the LO and TR directions (circles in
Fig.~\ref{Figure3-3}). We analyze our data by starting with the assumption that, in the low-$q$, low-$\omega$ limit, the spin fluctuations near an AFM
critical point can be described by a diffusive model \cite{Moriya1985,Inosov2009}. As in Ref. \cite{Diallo2010}, this model can be extended to include
anisotropic spin correlation lengths within the Fe layer,
\begin{eqnarray}
&&\chi^{\prime\prime}(\textbf{Q}_{\text{AFM}}+ \textbf{q}, \omega) \nonumber\\
&=& \frac{\chi_0\Gamma\omega}{\omega^2 + \Gamma^2[1 + \frac{(q_\texttt{x} + q_\texttt{y})^2}{2}\xi^2_{\texttt{LO}} +
\frac{(q_\texttt{x} - q_\texttt{y})^2}{2}\xi^2_{\texttt{TR}}]^2}.
\end{eqnarray}
Here $\chi_0$ represents the strength of the AFM response function, $\Gamma$ is a damping constant and $\xi_{\texttt{LO}}$ and $\xi_{\texttt{TR}}$ are spin
correlation lengths along LO and TR directions relative to $\textbf{Q}_{\texttt{AFM}}$. Fits to the neutron intensities,
$S(\textbf{Q},\omega)=f^{2}(|\textbf{Q}|)\chi^{\prime\prime}(\textbf{Q}_{\text{AFM}}+\textbf{q},\omega)(1-e^{-\hbar\omega/k_\texttt{B}T})^{-1}$, where
$f^{2}(|\textbf{Q}|)$ is the magnetic form factor of the Fe$^{2+}$ ion, were performed by a procedure where $\xi$ was varied while keeping $\Gamma$ = $9.5
\pm 1.0$ meV as obtained from fitting the 30 K data in Fig.~\ref{Figure4-4}(a). The resulting averaged anisotropy of the LO and TR spin-correlation lengths
is given by the ratio $\xi_{\texttt{LO}}$($10.4 \pm 0.6$ {\AA})/$\xi_{\texttt{TR}}$($5.9 \pm 0.4$ {\AA}) = $1.8 \pm 0.2$ at 4 K. Despite the expectation
that the diffusive model is valid only at low energies, we find that the parameters above describe constant energy cuts over the entire energy range in the
LO direction, as shown in Figs. 3(a)-3(e).  Along the TR direction, the diffusive model works well up to $\sim$80 meV [Figs. 3(c$'$)-3(e$'$)]. Above
$\sim$80 meV, the TR spectrum splits into two peaks as shown in Figs. 3(a$'$) and 3(b$'$) that cannot be described by the diffusive model (to be
discussed).

The spectral features within the diffusive model are described by a single peak centered at $\textbf{Q}_{\texttt{AFM}}$ whose $q$-space half-width at half
maximum (HWHM) depends on energy according to $q_{\texttt{HWHM}}^{2} = \xi^{-2}(\sqrt{2+\omega^{2}\Gamma^{-2}}-1)$. To confirm the validity of the
diffusive model over a wide energy range, we show that the Lorentzian HWHM of the constant energy cuts vs energy agrees with the $q_{\texttt{HWHM}}$ as
shown in Fig. 3(f), which also highlights the anisotropy between LO and TR widths.

The in-plane anisotropy is consistent with an itinerant description of the dynamical magnetic susceptibility due to the ellipticity of the electron pockets
at $\textbf{Q}_{\texttt{AFM}}$ \cite{Knolle2010}. Alternatively, a local-moment picture yields the same phenomenology and allows the anisotropy to be
associated with competing NN ($J_1$) and NNN ($J_2$) exchange interactions. We thus note that our analysis cannot address the question of the
appropriateness of an itinerant or local-moment description for the iron arsenides. For a region of stripe AFM correlations corresponding to
$\textbf{Q}_{\texttt{AFM}} = (\frac{1}{2} \frac{1}{2})$, the LO (TR) direction is along AFM (FM) correlated spins [Fig. 1(a)]. When evaluated in the
$J_1$-$J_2$ model, the dynamical magnetic susceptibility near $\textbf{Q}_{\texttt{AFM}}$ takes an identical form to model Eq. (1), allowing the anisotropy
observed in the correlation lengths to be associated with the exchange ratio. Specifically it can be shown that $\frac{J_1}{J_2}=2\frac{\xi^2_{\texttt{LO}}
- \xi^2_{\texttt{TR}}}{\xi^2_{\texttt{LO}} + \xi^2_{\texttt{TR}}}$ \cite{Diallo2010}. $\xi_{\texttt{TR}} < \xi_{\texttt{LO}}$ indicates that $J_1$ is AFM
and acts to destabilize (stabilize) FM (AFM) near-neighbor correlations. Using the experimental correlation lengths we obtain $\frac{J_1}{J_2}$ = 1.0 $\pm$
0.2 at 4 K, which is similar to the ratio found in Ba(Fe$_{0.935}$Co$_{0.065}$)$_2$As$_2$ \cite{Lester2010}, and in the AFM ordered CaFe$_2$As$_2$
\cite{Diallo2009,Zhao2009}, and indicates NNN interactions are important in the iron arsenides. We emphasize that the observed anisotropy does not break
the fourfold symmetry of the Fe square lattice but it does imply anisotropic correlations and hence interactions between NN spins. The inferred exchange
ratio is indeed within the regime of frustrated magnetism ($\frac{J_1}{J_2} < 2$) where spin nematic properties have been predicted.

The absence of any splitting in the LO direction is unusual because one expects the diffusive excitations to evolve into damped spin-wave modes at short
wavelengths (high energies). Assuming a typical conical spin-wave dispersion, constant energy cuts should display ringlike features at high energy, instead
of split maxima. Rather, the TR splitting bears some similarity to magnetic excitations in incommensurate spin-density-wave systems such as Cr
\cite{Fawcett1988}, and in the FeTe$_{1-x}$Se$_{x}$ pnictide superconductors \cite{Lumsden2009-2}. The phenomenological Sato-Maki function \cite{Sato1974}
has been used to describe the spin fluctuations in these incommensurate systems. However, in our case, there is no indication of incommensurability at low
energies and we thus treat the TR excitations as damped propagating modes by including a ballistic term in the diffusive model,
\begin{eqnarray}
\chi^{\prime\prime}(\textbf{Q}_{\text{AFM}}+ \textbf{q}, \omega) =
\frac{\chi_0\Gamma\omega}{\omega^2 + \Gamma^2(1 + q^2\xi_{\texttt{E}}^2 -\frac{\omega^2\xi_{\texttt{E}}^2}{c^2})^2},
\end{eqnarray}
where $c$ corresponds to the velocity of the propagating mode and $\xi_{\texttt{E}}$ is an effective TR correlation length. The TR splitting above $\sim$80
meV can be fit to Eq. (2), shown as solid lines in Figs. 3(a$'$) and 3(b$'$). This produces $\xi_{\texttt{E}}$ = $7.4\pm0.8$ {\AA} and \emph{c} =
$245\pm10$ meV{\AA}. The obtained $\xi_{\texttt{E}}$ is somewhat larger than $\xi_{\texttt{TR}}$ = $5.9\pm0.4$ {\AA} from Eq. (1). Despite the unusual
splitting, we find that $c$ is comparable to the spin-wave velocity in the TR direction in AFM ordered CaFe$_2$As$_2$ (300-350 meV\AA)
\cite{Diallo2009,Zhao2009}. Lester \emph{et al}.\cite{Lester2010} report a similar TR velocity (230 meV{\AA}) in paramagnetic
Ba(Fe$_{0.935}$Co$_{0.065}$)$_2$As$_2$.

To better determine the evolution of the TR splitting, we also fit the experimental TR spectra with a symmetric pair of Lorentzians, whose peak splitting
is plotted in Fig. 3(f$'$). According to the ballistic model, a diffusive (single-peaked) response is obtained below $\omega_\texttt{d} =
c\xi_{\texttt{E}}^{-1} \approx$ 33 meV in the TR direction. Above this energy, the ballistic model predicts split peaks at $q = \pm c^{-1}\sqrt{\omega^{2}
- c^2 \xi^{-2}}$, which only approaches a damped simple harmonic oscillator (DSHO) response at high energies (with splitting $q = \pm \omega/c$), as shown
in Fig. 3(f$'$). The actual observation of a splitting depends on the statistical quality of the data which, in our case, allows a clear observation of the
TR splitting only above $\sim$80 meV. However, the relatively low value of $\omega_{\texttt{d}}$ explains the agreement between the TR velocities obtained
from our ballistic model and DSHO model of Lester {\it et al.} \cite{Lester2010}. In the LO direction, the magnetic response appears diffusive at all
energies (subject to the statistical quality of the data), apparently due to a much higher LO velocity. The report of a LO dispersion with $c_{\texttt{LO}}
=$ 580 meV{\AA} based on DHSO model analysis is probably a lower bound. The finite correlation length of the system allows one to estimate that
$\omega_{\texttt{d}} = c_{\texttt{LO}}\xi_{\texttt{LO}}^{-1} > $ 56 meV suggesting that most if not all of the LO data analyzed in Ref. \cite{Lester2010}
is within the diffusive limit.

Despite the agreement of the diffusive/ballistic models along the LO and TR directions, the nature of the excitations in the full $[H, K]$ plane is still
unclear. The unidirectional nature of the split modes and the strong anisotropy cannot be easily modeled by Eq. (2), which highlights the anomalous nature
of the high-energy magnetic spectrum.

We now discuss the effects of superconductivity on the anisotropic spin excitations, which were examined using the triple-axis spectrometer HB-3. Figure
\ref{Figure4-4} shows the inelastic intensity at constant $\textbf{Q} = (\frac{1}{2} \frac{1}{2} 0)$ as a function of energy transfer at 5 and 30 K.
Spectral weight is pushed to higher energies as superconductivity develops. To obtain magnetic scattering only, we measured identical energy scans at a
point in reciprocal space \textbf{Q} = (0.683 0.183 0) [Fig.\ \ref{Figure4-4}(b)], where no appreciable temperature dependence is observed between 5 and 30
K. We subtracted their average from the data at ${\bf Q} = (\frac{1}{2} \frac{1}{2} 0)$ to obtain the background-subtracted magnetic scattering spectra
shown in Fig.~\ref{Figure4-4}(a). The spectra can be fit, respectively, to a single imaginary-pole-response function in the normal state (overdamped
response) and a damped harmonic oscillator in the SC state. The normal state relaxation rate at the critical wave-vector is found to be
$\Gamma_\texttt{n}=9.5\pm1.0$ meV. In the SC state the damping rate has decreased to $\Gamma_\texttt{s}=6.6\pm0.4$ meV and the resonance energy is
$\hbar\Omega \sim 8.3$ meV.

To examine the spatial correlations associated with the spin resonance, we measured the momentum dependence at 8 meV and 5 K [Fig.~\ref{Figure2-2}(a)]. A
quasielliptical feature around ${\bf Q} = (\frac{1}{2} \frac{1}{2} 0)$ was observed similar to the data below $\sim$80 meV. The anisotropic scattering
extends beyond the instrumental resolution ellipsoid shown in Fig.~\ref{Figure2-2}(a). At 30 K [Fig.~\ref{Figure2-2}(b)], the normal-state spin
fluctuations display a similar momentum space anisotropy. The LO and TR cuts through the resonance can also be well fit to the diffusive model yielding
peak positions (TR) and peak widths that are consistent with those data as shown in Figs.\ \ref{Figure3-3}(f) and \ref{Figure3-3}(f$'$). We conclude that
within errors there are no changes in the spatial correlations above and below $T_\texttt{C}$. It is predominantly the spectrum of magnetic excitations
that is modified by superconductivity.

To summarize, we have observed collective magnetic excitations in SC Ba(Fe$_{0.926}$Co$_{0.074}$)$_2$As$_2$ close to the $\textbf{Q}_{\texttt{AFM}}$
wave-vector of the parent BaFe$_2$As$_2$ compound. At low energies, the excitations have a pronounced in-plane anisotropy that can be associated with
frustrated versus \emph{satisfied} NN interactions in the parameter regime of dynamic nematic correlations. The spin resonance in the SC state is found to
have the same anisotropy. At energies above $\sim$80 meV, quasipropagating modes are observed experimentally along the TR direction, while modes in the LO
direction appear to have a very large energy scale.

This research is supported by the U.S. Department of Energy (USDOE), Office of Basic Energy Sciences, Division of Materials Sciences and Engineering.  Ames
Laboratory is operated for the USDOE by Iowa State University under Contract No. DE-AC02-07CH11358. Johns Hopkins Institute for Quantum Matter is supported
by the USDOE under Grant No. DE-FG02-08ER46544. Research at Oak Ridge National Laboratory's High Flux Isotope Reactor and Spallation Neutron Source is
sponsored by the Scientific User Facilities Division, Office of Basic Energy Sciences, DOE. We thank M. J. Loguillo for assistance with the ARCS
measurements. We acknowledge the expert technical assistance of Scott Spangler at JHU in machining the sample mount for HB-3 measurements.

\end{document}